\documentclass[prb,twocolumn,noshowpacs,superscriptaddress,longbibliography]{revtex4-1}
\usepackage{CJK}
\usepackage{bm,bbm}
\usepackage{graphicx}
\usepackage{color}
\usepackage{amsmath,amssymb,amsfonts,amsthm}
\usepackage[colorlinks=true,linkcolor=blue,anchorcolor=red,citecolor=blue,urlcolor=blue]{hyperref}

\begin{document}

\title{Effective Long-Range Pairing and Hopping in Topological Nanowires Weakly Coupled to $ s $-Wave Superconductors}

\author{Huaiqiang Wang}
\affiliation{National Laboratory of Solid State Microstructures and department of Physics, Nanjing University, Nanjing, 210093, China}
\affiliation{Collaborative Innovation Center of Advanced Microstructures, Nanjing University, Nanjing 210093, China}

\author{L. B. Shao}
\email[]{lbshao@nju.edu.cn}
\affiliation{National Laboratory of Solid State Microstructures and department of Physics, Nanjing University, Nanjing, 210093, China}
\affiliation{Collaborative Innovation Center of Advanced Microstructures, Nanjing University, Nanjing 210093, China}

\author{ Y. X. Zhao }
\email[]{zhaoyx@nju.edu.cn}
\affiliation{National Laboratory of Solid State Microstructures and department of Physics, Nanjing University, Nanjing, 210093, China}
\affiliation{Collaborative Innovation Center of Advanced Microstructures, Nanjing University, Nanjing 210093, China}

\author{ L. Sheng }
\affiliation{National Laboratory of Solid State Microstructures and department of Physics, Nanjing University, Nanjing, 210093, China}
\affiliation{Collaborative Innovation Center of Advanced Microstructures, Nanjing University, Nanjing 210093, China}

\author{ B. G. Wang }
\affiliation{National Laboratory of Solid State Microstructures and department of Physics, Nanjing University, Nanjing, 210093, China}
\affiliation{Collaborative Innovation Center of Advanced Microstructures, Nanjing University, Nanjing 210093, China}

\author{ D. Y. Xing}
\email[]{dyxing@nju.edu.cn}
\affiliation{National Laboratory of Solid State Microstructures and department of Physics, Nanjing University, Nanjing, 210093, China}
\affiliation{Collaborative Innovation Center of Advanced Microstructures, Nanjing University, Nanjing 210093, China}

\begin{abstract}
	In this Letter, we first formulate an effective theory, which generally captures long-range proximity effects of a surface system weakly coupled to an s-wave superconductor. The long-range proximity effects include both the emergent long-range pairing and hopping interactions in the surface system. We then model the Rashba spin-orbit-coupled nanowire in proximity with an s-wave superconductor by taking into account the emergent nonlocal effects in the weak-coupling limit. In this limit the induced superconducting pair potential is found much smaller than that of the host superconductor, which is in good agreement with recent experiments. Compared with the previously considered strong coupling limit with local proximity effects, the long-range interactions can significantly modify the topological phase diagram, and considerably lower the threshold magnetic field for the emergence of Majorana zero modes.
\end{abstract}
\maketitle

\textit{Introduction}--
Thanks to the development of nano-fabrication techniques, the hybrid structure of nanoscale material such as quantum dots and nanowires in proximity with superconductors can be readily realized experimentally~\cite{deon,hof,her,gram1}, where the superconducting proximity effect can lead to a diversity of subgap states in nanodevices, such as the Cooper-pair splitter~\cite{loss2,hof,her}, Andreev bound states~\cite{gram,ion,liu,tosi}, and Andreev tunneling \cite{oiw}. In particular, a one-dimensional (1D) topological superconductor can be realized by a  spin-orbit-coupled nanowire in the proximity with $s$-wave superconductors~\cite{alicea2,fu1,benak,qi1,das1,alicea}, and tremendous experimental efforts have been paid along this direction to confirm the existence of Majorana zero end-modes~\cite{zuo,deng,lisc,zhang1}. This is mainly due to that quasiparticle excitations of Majorana fermions located at the defects of topological superconductor have been considered to be one of the most promising candidates for topological quantum computing hardware based on their nonabelian statistics~\cite{ivan,kitaev,kitaev1,fu1}.

All these researches underlined the importance of the superconducting proximity effect in understanding and exploring the novel quantum effects in hybrid nanostructures. Most of previous works concerning the proximity effect adopted the scenario that the induced superconducting pairing potential (SPP) is approximately equal to that of the host superconductor, which can be derived from strong couplings between the surface system and the host superconductor~\cite{james,lee, ali1}. However, experimental evidences showed that the induced SPP can be much smaller than that of the host superconductor~\cite{deng,Das2012}. This suggests that the strong coupling limit may not be the answer in some experiments, and therefore it is meaningful to explore other possibilities.

In this Letter, we establish an effective theory for the proximity effect by taking the weak-coupling limit, which gives arise small SPPs in good agreement with the experimental evidences. Furthermore, the resultant interactions feature long-range pairings and hoppings. It is worth to note that long-range pairings and hoppings of various forms in topological superconductors have recently attracted much attention~\cite{Niu,vodola,fu2,lepori,Alecce,Viyuela,Gong2016,Vodola2016Long,DeGottardi, Patrick}. However, before the present work, a microscopic theory for the emergence of these long-range interactions, to the best of our knowledge, was still absent, and therefore the exact form of the induced long-range interactions had not been determined. In our theory, both pairing and hopping are rigorously derived as a result from the crossed Andreev reflection, and exhibit exponentially damped oscillation with the characteristic length equal to the superconducting coherence length. Since the coherence length is typically much larger than the lattice constant, the nonlocal pairing and hopping virtually correspond to long-range proximity effects.


We then apply the uncovered long-range interactions to study a 1D Rashba nanowire in proximity with an $s$-wave superconductor, whose importance has been addressed above.  Compared with the previously considered strong-coupling limit with local proximity effects~\cite{alicea}, the long-range interactions from the weak-coupling limit have several significant experimental consequences: i) The topological phase diagram has been dramatically changed, where particularly the topological regions are no longer symmetric under the inversion of the onsite energy; ii) As a result of the deformed phase diagram, a much smaller critical magnetic field is required for the emergence of Majorana zero modes by tuning the gate voltage of nanowire;  iii) Topological phase transitions explicitly depend on the Fermi momentum of the host superconductor. Moreover, it is worth to note that the framework established in our work can be applied to various other surface-superconductor hybrid systems in the weak-coupling case as well.

\textit{The effective theory--}
Let us start with a binary system consisting of two adatoms A and B in proximity with an $ s $-wave superconductor. Then, the Hamiltonian can be written as
\begin{equation}\label{model}
\hat{H}=\hat{H}_{\mathrm{N}}+\hat{H}_{\mathrm{SC}}+\hat{H}_{\mathrm{T}},
\end{equation}
where
\begin{align*}
&\hat{H}_{\mathrm{N}}=\sum_{i,\sigma}\left(\varepsilon_0-\mu_F\right)c_{i\sigma}^\dagger c_{i\sigma},\\
&\hat{H}_{\mathrm{SC}}=\int d^3r~\Psi^{\dagger}(\mathbf{r})h_{\mathrm{SC}}\Psi(\mathbf{r}),\\
&\hat{H}_\mathrm{T}=t_0\int d^3r~\eta^\dagger(\mathbf{r})\Psi(\mathbf{r})+\mathrm{H.c.}.
\end{align*}
Here, 
$\hat{H}_{\text{N}}$ is the Hamiltonian of the attached system, where $i=A$ or $B$ labeling the two adatoms, and the on-site energy $\varepsilon_0$ can be controlled by the gate voltage.
$\hat{H}_{\text{SC}} $ is the Hamiltonian of the host superconductor with the Nambu spinor $ \Psi(\mathbf{r})= [\psi_{\uparrow}(\mathbf{r}),\psi_{\downarrow}^{\dagger}(\mathbf{r})]^T $. The Bogoliubov-de Gennes (BdG) Hamiltonian is given by $ h_{\text{SC}}=[\hat{\mathbf{p}}^2/(2m)-\mu_{F}]\tau_3+\Delta_0\tau_1 $, 
where $ \mu_F $ and $ \Delta_0 $ are the chemical potential and SPP, respectively.  $\hat{H}_{\text{T}}$ describes the coupling between the surface system and host superconductor, 
where $ \eta(\mathbf{r})=[d_\uparrow(\mathbf{r}),-d_\downarrow^\dagger(\mathbf{r})]^T $, with $ d_{\sigma} =c_{A\sigma}\delta(\mathbf{r}-\mathbf{r}_A)+c_{B\sigma}\delta(\mathbf{r}-\mathbf{r}_B) $.

In contrast to previous works, we consider the weak coupling limit,
\begin{equation}
t_0, \ \left| \varepsilon_0-\mu_F \right|\ll \Delta_0,\label{cond}
\end{equation}
namely, that the characteristic energy scale of surface system is much smaller than  $ \Delta_0 $, so is the coupling strength between the surface system and host superconductor. This condition corresponds to physics in the subgap region, since the excitations of the host superconductor have much higher energies than those of the attached system. In fact, there are experimental evidences favoring the weak-coupling limit, rather than the strong-coupling limit. For the later case, the two adatoms together with the superconductor are tightly bounded, approximately with the same SPP, $\Delta_{0}$~\cite{lee}. But, this contradicts with the experimental evidences in Refs.~\cite{deng} and \cite{Das2012}, which showed that the induced SPP is much smaller than that of the host superconductor. For instance, in Ref.~\cite{deng}, the SPP of the host superconductor Nb is about $ 1.55$ meV, much larger than the induced SPP of InSb, which is about $ 0.25$ meV. As we shall see in Eq.~\eqref{effective-interactions}, the weak coupling limit can lead to small SPP in agreement with these experiments.

Different from the renormalization method in the strong-coupling limit~\cite{lee}, the effective action for the weak-coupling limit can be derived by integrating out the fermionic degree of freedom, $\Psi(\mathbf{r})$, of the superconductor, which gives
\begin{equation}
 S_{\mathrm{eff}}=t_0^2\int d^4x d^4x'~\eta^{\dagger}(x)G(x-x')\eta(x')\label{eq:seff}
\end{equation}
with $x^\mu=(\mathbf{r},\tau)$.
The Fourier transform of $G(x-x')$ is just the Matsubara Green's function,
$
G\left(i\omega_{n},\mathbf{k}\right)=1/[i\omega_{n}-h_{\mathrm{sc}}(\mathbf{k})]
$,
and $ h_{\mathrm{sc}}(\mathbf{k}) $ can be diagonalized by a unitary transformation, 
$
\varepsilon_{\mathbf{k}}\tau_{3}=U_{\mathbf{k}}h_{\mathrm{sc}}(\mathbf{k}) U_{\mathbf{k}}^{\dagger}
$,
where $ \varepsilon_{\mathbf{k}}=\sqrt{\xi(k)^2+\Delta_{0}^{2}} $ with $ \xi(k)={\hbar^2 k^2}/{2m}-\mu_{F} $. Then, by employing the contour integral trick for the summation over the Matsubara frequencies, we can explicitly calculate $G(\tau-\tau',\mathbf{k})$ as
\begin{equation}
\label{sum}
\begin{split}
&T\sum_{i\omega_{n}}U_{\mathbf{k}}^{\dagger}\frac{1}{i\omega_{n}-\varepsilon_{\mathbf{k}}\tau_{3}}U_{\mathbf{k}}e^{-i\omega_{n}(\tau-\tau')}\\
=&U_{\mathbf{k}}^{\dagger}\left(\begin{smallmatrix}
-\Theta\left(\tau-\tau'\right) & 0 \\ 0 &\Theta\left(\tau'-\tau\right)
\end{smallmatrix}\right)U_{\mathbf{k}}~ e^{-\varepsilon_{\mathbf{k}}\left|\tau-\tau'\right|},
\end{split}
\end{equation}
where $ T $ is the temperature, and $ \Theta(\tau) $ is the step function. Hence, the imaginary-time correlation length is $1/\varepsilon_{\mathbf{k}}\sim1/\Delta_0$, which is very small, because $ \Delta_{0} $ is large in the weak coupling limit. This actually reflects the uncertainty principle, since the time uncertainty can be evaluated as $1/\Delta_0$. As a result, at low temperatures and time scale much larger than $1/\Delta_0$, we can approximate that $G(x-x')\approx-\delta(\tau-\tau')\mathcal{K}(\mathbf{r}-\mathbf{r'})$, where  $\mathcal{K}(\mathbf{r})$ is the Fourier transform of $1/h_{sc}(\mathbf{k})$. In other words, the imaginary-time correlation is effectively instantaneous in the weak-coupling limit, and accordingly, the effective Hamiltonian reads
\begin{equation}
\hat{H}_{\mathrm{eff}}=-t_0^2\int d^3rd^3r'~\eta^{\dagger}(\mathbf{r})\mathcal{K}(\mathbf{r}-\mathbf{r'})\eta(\mathbf{r'}).\label{heff1}
\end{equation} 
The result can also be understood as following. As a result of the weak coupling limit in Eq.~\eqref{cond}, only the frequencies $ |i\omega_{n}|\sim\epsilon-\mu_F\ll\Delta_{0} $ are relevant for the low-energy effective theory. Because the energy scale of $ h_{\mathrm{sc}}(\mathbf{k})$ is about $\Delta_{0} $,  the Green's function can be approximated as $ G\left(i\omega_{n},\mathbf{k}\right)\approx-1/h_{\mathrm{sc}}(\mathbf{k}) $, from which Eq.~\eqref{heff1} can be derived.

The effective Hamiltonian of Eq.~\eqref{heff1} can be derived explicitly. Since $h_{SC}(\mathbf{k})$ is isotropic, $\mathcal{K}(\mathbf{r}-\mathbf{r}')$ depends only on the distance $l=|\mathbf{r}-\mathbf{r'}|$. Then, we can proceed straightforwardly to obtain
$
\mathcal{K}(l)=\int d\xi N(0)~1/h_{SC}[\sin(kl)/(kl)],
$
where $k=k_{F}(1+{\xi}/{\mu_{F}})^{1/2}$ with $ k_F $ the Fermi momentum, and $N(0)$ is the density of states near the Fermi energy. Since the main contribution comes from the states near the Fermi energy with $\xi\ll\mu_{F} $, we can approximate that $k\approx k_{F}[1+\xi/(2\mu_{F})]$, and extend the integration interval of $\xi$ to be $\left(-\infty,\infty\right)$, which gives
\begin{equation}
\mathcal{K}(l)=\pi N(0)\frac{\tau_{3}\cos k_{F}l+\tau_{1}\sin k_{F}l}{k_{F}l}e^{-{l}/{\xi_{0}}}. \label{K-l}
\end{equation}
Finally, substituting Eq.~\eqref{K-l} into Eq.~\eqref{heff1}, and recalling the explicit expression of $\eta(\mathbf{r})$, we obtain the effective Hamiltonian of the attached system consisting of adatoms $ A $ and $ B $, which is given by
\begin{equation}
\label{eff}
\hat{H}_{\mathrm{eff}}=\widetilde{\Delta}(c_{A\uparrow}^\dagger c_{B\downarrow}^\dagger+c_{B\uparrow}^\dagger c_{A\downarrow}^\dagger)-\widetilde{t}\sum_{\sigma}c_{A\sigma}^\dagger c_{B\sigma}+\mathrm{H.c.}.
\end{equation}
$\widetilde{t}$ and $\widetilde{\Delta}$ depend on the distance $l$ between the two adatoms as
\begin{equation}
\widetilde{t}(l)+i\widetilde{\Delta}(l)=[\widetilde{\Delta}_0/(k_Fl)]\exp(-l/{\xi_0}+ik_Fl), \label{effective-interactions}
\end{equation}
where the on-site SPP is given by $\widetilde{\Delta}_0=\pi t_{0}^{2}N(0)$, and the coherence length of the host superconductor $ \xi_{0}={2\mu_{F}}/(k_{F}\Delta_0) $. 

Several comments are ready for the result of Eq.~\eqref{eff}. First, we emphasize that, in contrast to the strong coupling case, the induced SPP in the weak coupling limit of Eq.~\eqref{cond} is much smaller than the SPP of the host superconductor, namely, that $\widetilde{\Delta}(l)\ll \Delta_0$. As afore claimed, this is consistent with recent experimental evidences~\cite{deng,Das2012}. Furthermore, $\widetilde{\Delta}_0$ is even independent of $\Delta_{0}$, which resembles the result in quantum-dot systems~\cite{james}. Second, we notice that both terms in Eq.~\eqref{eff} arise from the crossed Andreev reflection, where two electrons with opposite spins transfer from adatom A and B, respectively, into the superconductor, and then form a Cooper pair, or vice versa. Third, $\widetilde{\Delta}_{AB}$ and $\widetilde{t}_{AB}$ spatially oscillate with the atomic length $k_F^{-1}$ of the superconductor, and exponentially decay with the decaying length $ \sim\xi_{0} $. However, they are still long-range interactions, since $ \xi_{0}$ is typically much larger than atomic spacings.

\textit{An application to nanowires--}
As an application of the effective long-range hopping and pairing above, we consider a 1D Rashba spin-orbit-coupled nanowire in proximity with an $s$-wave superconductor, which has been widely used for realizing Majorana zero modes~\cite{alicea,Lutchyn,Oreg}. The Hamiltonian of an isolated 1D nanowire with Rashba spin-orbit interaction in the presence of a parallel magnetic field is given by
\begin{multline}
\label{h0}
\hat{H}_{\mathrm{0}}=\sum_{i}(-tc_{i}^{\dagger}c_{i+1}+\mathrm{H.c.})+\sum_{i}\bar{\epsilon} c_{i}^{\dagger}c_{i} \\
+\sum_{i}(i\lambda_{R}c^{\dagger}_{i}\sigma_{2}c_{i+1}+\mathrm{H.c.})+\sum_{i}h_{Z}c_{i}^{\dagger}\sigma_{1}c_{i}.
\end{multline}
The first term is a nearest neighbor hopping term, and the second term is an on-site energy term, where we have suppressed the spin index on the electron operators. The third term is a nearest neighbor Rashba term, and the fourth term corresponds to the Zeeman effect under the parallel magnetic field, where $\sigma_{i}$ with $i=1,2,3$ are the Pauli matrices.

By taking into account the long-range pairing and hopping terms from the proximity effect, the total Hamiltonian can be written as $ \hat{H}_{\mathrm{tot}}=\hat{H}_{\mathrm{0}}+\hat{H}_{\mathrm{eff}} $, where
\begin{equation}\label{heff}
\hat{H}_{\mathrm{eff}}=-\sum_{i,r\neq0}\widetilde{t}(r)c_{i}^{\dagger}c_{i+r}+\sum_{i,r}\widetilde{\Delta}(r)c_{i,\uparrow}^{\dagger}c^{\dagger}_{i+r,\downarrow}+\mathrm{H.c.}.
\end{equation}
The coefficients $\widetilde{t}(r)$ and $\widetilde{\Delta}(r)$ are given by Eq.~\eqref{effective-interactions}, where the integer $r$ corresponds to the distance between two atoms related by the long-range interactions. For the first term, the case with $r=0$ can be absorbed into the second term of Eq.~\eqref{h0} after lattice regularization, and therefore has been excluded in the summation. 
Accordingly, the BdG Hamiltonian is 
\begin{multline}
\label{hk}
h(k)=-[2t\cos k-\bar{\epsilon}+\widetilde{t}(k)]\tau_{3}\\
+\widetilde{\Delta}(k)\tau_{1}+2\lambda_{R}\sin k~\tau_{3}\sigma_{2}+h_{Z}\sigma_{1},
\end{multline}
with the Nambu spinor $(c_{k\uparrow},c_{k\downarrow},c^{\dagger}_{-k\downarrow},-c_{-k\uparrow}^{\dagger})^{T}$. $\tau_{i}$ with $i=1,2,3$ are the Pauli matrices acting in the particle-hole space. The explicit expressions of $\widetilde{t}(k)$ and $\widetilde{\Delta}(k)$ are given by
\begin{equation}
\label{tdk}
[\widetilde{t}(k)+i\widetilde{\Delta}(k)]/\widetilde{\Delta}_0=
i+\sum_{r=1}^{\infty}\frac{2\cos kr}{k_{F}r}e^{-{r}/{\xi_{0}}+ik_{F}r}.
\end{equation}
The dependences of $\widetilde{t}(k=0/\pi)$ and $\widetilde{\Delta}(k=0/\pi)$ on $k_F$ and $\xi_0$ are plotted in the Supplemental Material (SM)~\cite{Supp}.
\begin{figure}
	\centering
	\includegraphics[scale=0.8]{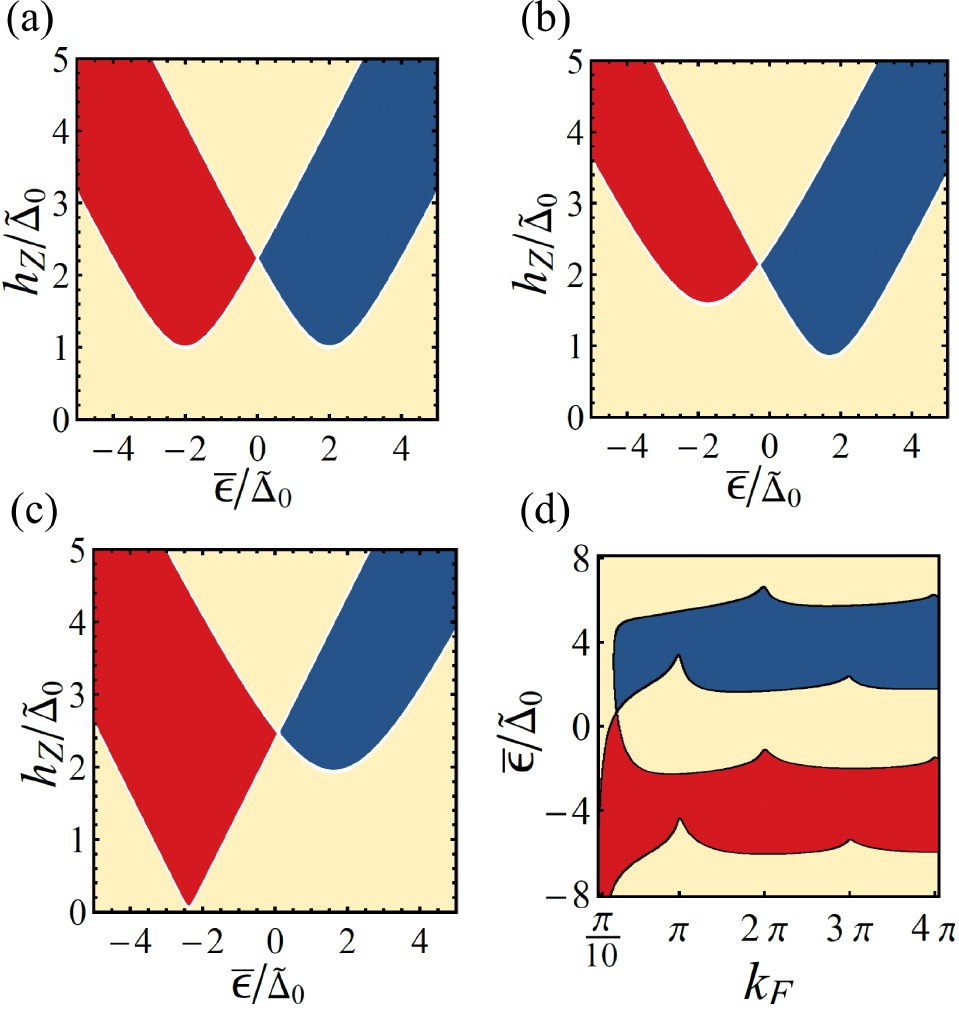}
	\caption{The $h_{Z}$-$\bar{\epsilon}$ phase diagram of the nanowire system  in the absence of long-range interactions (a), and in the presence of long-range interactions for $k_{F}=1.2\pi$ (b) and $0.5\pi$ (c), and $\bar{\epsilon}$-$k_{F}$ phase diagram with $h_{Z}=4$ (d).  Here $W=-1$ $(+1)$ for the red-color (blue-color) region and $W=0$ for the yellow-color region. The other parameters are taken as $t/\widetilde{\Delta}_{0}=1$ and $\xi_{0}=10$. }\label{diag}
\end{figure}

\textit{Topological phases--} We proceed to discuss the topological invariant of the system. The nanowire with long-range interactions belongs to class D in the Altland-Zirnbauer symmetry classification~\cite{Schn}, where particle-hole symmetry for Eq.~\eqref{hk} is represented by $\mathcal{C}=\tau_2\sigma_2 K$ with $K$ the complex conjugate. 1D topological superconductors in class D have a $\mathbb{Z}_2$ topological classification, and the corresponding topological invariant is just the quantized Berry phase of valence bands in unit of $\pi$ mod $2$~\cite{Schn,kitaev1}.  For Eq.~\eqref{hk}, the Berry phase is given explicitly by

\begin{equation}
W=[\mathrm{sgn}Z(\pi)-\mathrm{sgn}Z(0)]/2,\label{pt}
\end{equation}
where $Z(k)=(\widetilde{\Delta}(k)+2i\lambda_{R}\sin k)^2-h_{Z}^{2}+[2t\cos k-\bar{\epsilon}+\widetilde{t}(k)]^{2}$, and in particular,
$Z(0/\pi)=[2t+\widetilde{t}(0/\pi)-\bar{\epsilon}]^{2}-[h_{Z}^{2}-\widetilde{\Delta}^{2}(0/\pi)]$~\cite{Tewa2012, Waka2014, Wang2017}. A detailed derivation  can be found in the SM~\cite{Supp}.
The topological phase diagram of the nanowire system can be determined by Eq.~\eqref{pt}, and the phase diagrams are shown in Fig.~\ref{diag}, where we have chosen $t=\widetilde{\Delta}_{0}=1$ and $\lambda_{R}>0$ without loss of generality. In the SM~\cite{Supp}, we also numerically solve the Majorana zero modes at the ends of the nanowire, which is consistent with the bulk topological invariant.

Compared with the strong-coupling limit, the weak-coupling limit with the long-range interactions leads to three significant differences. First, the symmetry of the phase diagram under $\bar{\epsilon}$ to $-\bar{\epsilon}$ is violated by the long-range interactions.   
Taking $\xi_{0}\rightarrow 0$, the model with the strong coupling limit in Ref.~\cite{alicea} is formally recovered with $\widetilde{t}(k)=0$ and $\widetilde{\Delta}(k)=\widetilde{\Delta}_{0}$ from Eq.~\eqref{tdk}. We can derive the phase diagram of Fig.~\ref{diag}(a) from Eq.~\eqref{pt}, which is
mirror symmetric with respect to $ \bar{\epsilon}=0 $. However, the symmetry of phase diagram is broken, as observed in Figs.~\ref{diag}(b) and (c), when the long-range hopping and pairing are added. This can be understood by the following argument. In the case of $\widetilde{t}(k)=0$ and $\widetilde{\Delta}(k)=\widetilde{\Delta}_{0}$,  by taking $ \bar{\epsilon}\rightarrow-\bar{\epsilon} $ and $ k\rightarrow k+\pi $ in Eq.~\eqref{hk}, $Z(k)$ is transformed to $Z^{*}(k)$, which inverses the sign of the winding number $W$ of Eq.~\eqref{pt}. But, this symmetry is not intrinsic to the system, and can be violated when the long-range interactions arise. Particularly, $\widetilde{t}(0)$ and $\widetilde{t}(\pi)$ [$\widetilde{\Delta}(0)$ and $\widetilde{\Delta}(\pi)$] are in general unequal in $Z(0/\pi)$, leading to different sizes of the blue-colored and red-colored regions in Fig.~\ref{diag}(b) and (c).

Second, as a result of the asymmetric phase diagram, the threshold magnetic field at about $\bar{\epsilon}=-2.5 $ in Fig.~\ref{diag}(c)   for the topologically nontrivial phase  is vanishingly small  compared to the effective pairing potential $ \widetilde{\Delta}_0 $. This indicates an enlarged topological nontrivial region (with red color)  in the presence of the long-range hopping and pairing. In the strong coupling case, it was proposed that the condition $ h_Z>\Delta_{0} $ should be required for experimental realization of Majorana zero modes~\cite{alicea}. But, in the weak coupling case neglecting the long-range interactions, as shown in Fig.~\ref{diag}(a), the condition above has been replaced with a relaxed condition $ h_Z>\widetilde{\Delta}_{0} $, with $\widetilde{\Delta}_{0} $ being significantly smaller than  $ \Delta_{0} $.  This condition is further relaxed when the long-range interactions are included. The threshold magnetic field can even be much lower than $\widetilde{\Delta}_{0} $, as seen in Fig.~\ref{diag}(c). 

Third, in the weak-coupling limit, changing $k_F$ leads to topological phase transitions as shown in Fig.~\ref{diag}(d).  In contrast, the phase diagram in the strong coupling limit has no explicit $k_F$ dependence.  Figure~\ref{diag}(d) shows the $\bar{\epsilon}$-$k_{F}$ phase diagram with  $h_{Z}=4$, where topological phase transitions occur by varying $k_{F}$ at given on-site energies $\bar{\epsilon}$. This is because $\widetilde{t}(k)$ and $\widetilde{\Delta}(k)$ oscillate with $k_{F}$, as seen from Eq.~\eqref{tdk}.



\textit{Summary--}
In summary, we have rigorously derived the effective Hamiltonian for a surface system weakly coupled to an $ s $-wave superconductor. The induced long-range pairing and hopping are much smaller than the SPP of the host superconductor, in consistent with recent experiments. They exhibit exponentially damped oscillation with the characteristic length equal to the coherence length of host superconductor. Taking into account the long-range effects, we studied a Rashba spin-orbit-coupled nanowire in proximity with an $ s $-wave superconductor. Several experimental consequences on the topological phase diagram have been discussed. 
Finally, we note that the established framework is applicable to other hybrid systems of superconductors, where weak coupling takes place.

\begin{acknowledgments}
	\textit{Acknowledgments--}This work was supported by the National Key P$\&$D Program of China under Grants No. 2017YFA0303203, No. 2015CB921202, and No. 2014CB921103, and the National Natural Science Foundation of China under Grants No. 11704180.
\end{acknowledgments}

\bibliography{bibfile}

\widetext
\clearpage
\begin{center}
	\textbf{\large Supplemental Materials:\\ ``Effective Long-Range Pairing and Hopping in a Surface System Weakly Coupled to an $ s $-Wave Superconductor"}
\end{center}
\setcounter{equation}{0}
\setcounter{figure}{0}
\setcounter{table}{0}
\setcounter{page}{1}
\makeatletter
\renewcommand{\theequation}{S\arabic{equation}}
\renewcommand{\thefigure}{S\arabic{figure}}
\renewcommand{\bibnumfmt}[1]{[S#1]}
\renewcommand{\citenumfont}[1]{S#1}



\section{Derivation of Effective Hamiltonian}
In this section, we present technical details for the derivation of the effective theory $S_{eff}$ in the main text.
After integrating out the fermionic fields of the host superconductor, the effective action is given as
\begin{equation}\label{eff-action}
S_{\mathrm{eff}}=t_0^2\int d^4xd^4x'\eta^\dagger(x)G(x-x')\eta(x'),
\end{equation}
where $ x=(\tau,\mathbf{r}) $ and  
\begin{equation}
G(x-x')=T\sum_{i\omega_n}\int\frac{d^3k}{(2\pi)^3}G(i\omega_n,\mathbf{k})e^{i\mathbf{k}\cdot\left(\mathbf{r}-\mathbf{r}'\right)-i\omega_{n}(\tau-\tau')}.
\end{equation}
The Green's function $ G(i\omega_n,\mathbf{k})=1/\left(i\omega_n-h_{\mathrm{SC}}(\mathbf{k})\right) $ with $\omega_{n}$ the Matsubara frequencies. We can diagonalize the Hamiltonian of host superconductor by a unitary matrix $ U_{\mathbf{k}} $ as
\begin{equation}\label{diag}
\epsilon_{\mathbf{k}}\tau_3=U_{\mathbf{k}}h_{\mathrm{SC}}(\mathbf{k})U_{\mathbf{k}}^{\dagger},
\end{equation}
where $ \epsilon_{\mathbf{k}}=\sqrt{\xi(k)^2+\Delta_0^2} $ with $ \xi(k)=\hbar^2k^2/2m-\mu_F $.
Then, the Green's function $ G(i\omega_n,\mathbf{k}) $ can also be diagonalized by $ U_{\mathbf{k}} $,
\begin{equation}
G(i\omega_n,\mathbf{k})=U_{\mathbf{k}}^{\dagger}\begin{pmatrix}
\frac{1}{i\omega_n-\epsilon_{\mathbf{k}} }& 0 \\ 0 & \frac{1}{i\omega_n+\epsilon_{\mathbf{k}} }
\end{pmatrix}U_{\mathbf{k}}.
\end{equation} 
The summation of Matsubara frequency can be derived by contour integral trick and the Green's function in real space and imaginary time can be written as
\begin{equation}
\label{green-real}
\begin{split}
G(x-x')=\int\frac{d^3\mathbf{k}}{(2\pi)^3}U_{\mathbf{k}}^{\dagger}\left(\begin{smallmatrix}
-\Theta\left(\tau-\tau'\right) & 0 \\ 0 &\Theta\left(\tau'-\tau\right)
\end{smallmatrix}\right)U_{\mathbf{k}}~ e^{-\varepsilon_{\mathbf{k}}\left|\tau-\tau'\right|}~e^{i\mathbf{k}\cdot\left(\mathbf{r}-\mathbf{r}'\right)},
\end{split}
\end{equation}
where $ \Theta\left(\tau-\tau'\right) $ is the step function. As discussed in the main text, because of the factor $ e^{-\varepsilon_{\mathbf{k}}\left|\tau-\tau'\right|} $ in Eq.~\eqref{green-real}, the time correlation length is about $ 1/\varepsilon_{\mathbf{k}}\sim1/\Delta_0 $, which is very small in the weak coupling limit. As a result, the interaction of effective action in Eq.~\eqref{eff-action} is approximately instantaneous. Substituting Eq.~\eqref{green-real} into Eq.~\eqref{eff-action}, we have
\begin{equation}
\label{eff-action1}
\begin{split}
S_{\mathrm{eff}}&=t_0^2\int d^4xd^4x'\int\frac{d^3k}{(2\pi)^3}~\eta^\dagger(\tau,\mathbf{r})U_{\mathbf{k}}^{\dagger}\left(\begin{smallmatrix}
-\Theta\left(\tau-\tau'\right) & 0 \\ 0 &\Theta\left(\tau'-\tau\right)
\end{smallmatrix}\right)U_{\mathbf{k}}~\eta(\tau',\mathbf{r}')~ e^{-\varepsilon_{\mathbf{k}}\left|\tau-\tau'\right|}~e^{i\mathbf{k}\cdot\left(\mathbf{r}-\mathbf{r}'\right)}\\
&\approx t_0^2\int d^4xd^4x'\int\frac{d^3k}{(2\pi)^3}~\eta^\dagger(\tau,\mathbf{r})U_{\mathbf{k}}^{\dagger}\left(\begin{smallmatrix}
-\Theta\left(\tau-\tau'\right) & 0 \\ 0 &\Theta\left(\tau'-\tau\right)
\end{smallmatrix}\right)U_{\mathbf{k}}~\eta(\tau,\mathbf{r}')~ e^{-\varepsilon_{\mathbf{k}}\left|\tau-\tau'\right|}~e^{i\mathbf{k}\cdot\left(\mathbf{r}-\mathbf{r}'\right)}\\
&=t_0^2\int d\tau\int d^3rd^3r'\int\frac{d^3k}{(2\pi)^3}~\eta^\dagger(\tau,\mathbf{r})U_{\mathbf{k}}^{\dagger}\left(\begin{smallmatrix}
-\frac{1}{\varepsilon_{\mathbf{k}}} & 0 \\ 0 & \frac{1}{\varepsilon_{\mathbf{k}}}
\end{smallmatrix}\right)U_{\mathbf{k}}~\eta(\tau,\mathbf{r}')~e^{i\mathbf{k}\cdot\left(\mathbf{r}-\mathbf{r}'\right)}\\
&=-t_0^2\int d\tau\int d^3rd^3r'\int\frac{d^3k}{(2\pi)^3}~\eta^\dagger(\tau,\mathbf{r})\frac{1}{h_{\mathrm{SC}}(\mathbf{k})}~\eta(\tau,\mathbf{r}')~e^{i\mathbf{k}\cdot\left(\mathbf{r}-\mathbf{r}'\right)}.
\end{split}
\end{equation}
In the second line, we have used $ \eta(\tau',\mathbf{r}')\approx\eta(\tau,\mathbf{r}') $, since time correlation length is very small as being discussed in the main text. In the third line, we integrated out $ \tau' $. The fourth equality was derived by using Eq.~\eqref{diag}. From the effective action in Eq.~\eqref{eff-action1}, the effective Hamiltonian is given by
\begin{equation}
\hat{H}_{\mathrm{eff}}=-t_0^2\int d^3rd^3r'~\eta^{\dagger}(\mathbf{r})\mathcal{K}(\mathbf{r}-\mathbf{r'})\eta(\mathbf{r'}),\label{heff1}
\end{equation}
where $ \mathcal{K}(\mathbf{r}-\mathbf{r'}) $ is 
\begin{equation}
\label{KK}
\mathcal{K}(\mathbf{r}-\mathbf{r'})=\int\frac{d^3k}{(2\pi)^3}\frac{1}{h_{\mathrm{SC}}(\mathbf{k})}e^{i\mathbf{k}\cdot\left(\mathbf{r}-\mathbf{r}'\right)}.
\end{equation}
Since $h_{SC}(\mathbf{k})$ has rotational symmetry, $ \mathcal{K}(\mathbf{r}-\mathbf{r'}) $ in Eq.~\eqref{KK}  depends only on the distance $ l=\left|\mathbf{r}-\mathbf{r'} \right|  $, namely, that
\begin{equation}
\begin{split}
\mathcal{K}(l)&\approx N(0)\int d\xi \int\frac{\sin\theta d\theta d\varphi}{4\pi}~\frac{\xi\tau_{3}+\Delta_{0}\tau_{1}}{\xi^{2}+\Delta_{0}^{2}}e^{ikl\cos\theta}\\
&=N(0)\int d\xi \frac{\xi\tau_{3}+\Delta_{0}\tau_{1}}{\xi^{2}+\Delta_{0}^{2}}\frac{\sin(kl)}{kl},
\end{split}
\label{eq:green}
\end{equation}
where $ N(0) $ is the density of states of the host superconductor near the Fermi energy.   Because the main excitations of the host superconductor relevant to the effective Hamiltonian is near the Fermi energy, we have $ \xi\ll\mu_F $ and $ k\approx k_F[1+{\xi}/(2\mu_F)] $. Hence, we can evaluate Eq.~\eqref{eq:green} by extending the integration domain from $ (-\omega_D,\omega_D) $ to $ (-\infty,\infty) $ with $ \omega_D $ being the Debye frequency of the host superconductor. 
Although the diagonal term of $\mathcal{K}(l=0)$ becomes divergent for $ \xi\in(-\infty,\infty) $, it can be absorbed into the renormalized chemical potential, which is determined by equilibrium with the host superconductor. 
Consequently, 
only the off-diagonal terms corresponding to  an induced $ s $-wave pairing potential are physically interesting.


\section{The Emergent pairing and hopping in a nanowire}

In the main text, the emergent pairing and hopping terms in a nanowire are given by
\begin{eqnarray}
\widetilde{t}(k)/\widetilde{\Delta}_0&=&\sum_{r=1}^{\infty}\frac{2}{k_{F}r}e^{-\frac{r}{\xi_{0}}}\cos k_{F}r \cos kr,\\
\widetilde{\Delta}(k)/\widetilde{\Delta}_0&=&1+\sum_{r=1}^{\infty}\frac{2}{k_{F}r}e^{-\frac{r}{\xi_{0}}}\sin k_{F}r \cos kr.
\end{eqnarray}
\begin{figure}[h]
	\centering
	\includegraphics[scale=0.5]{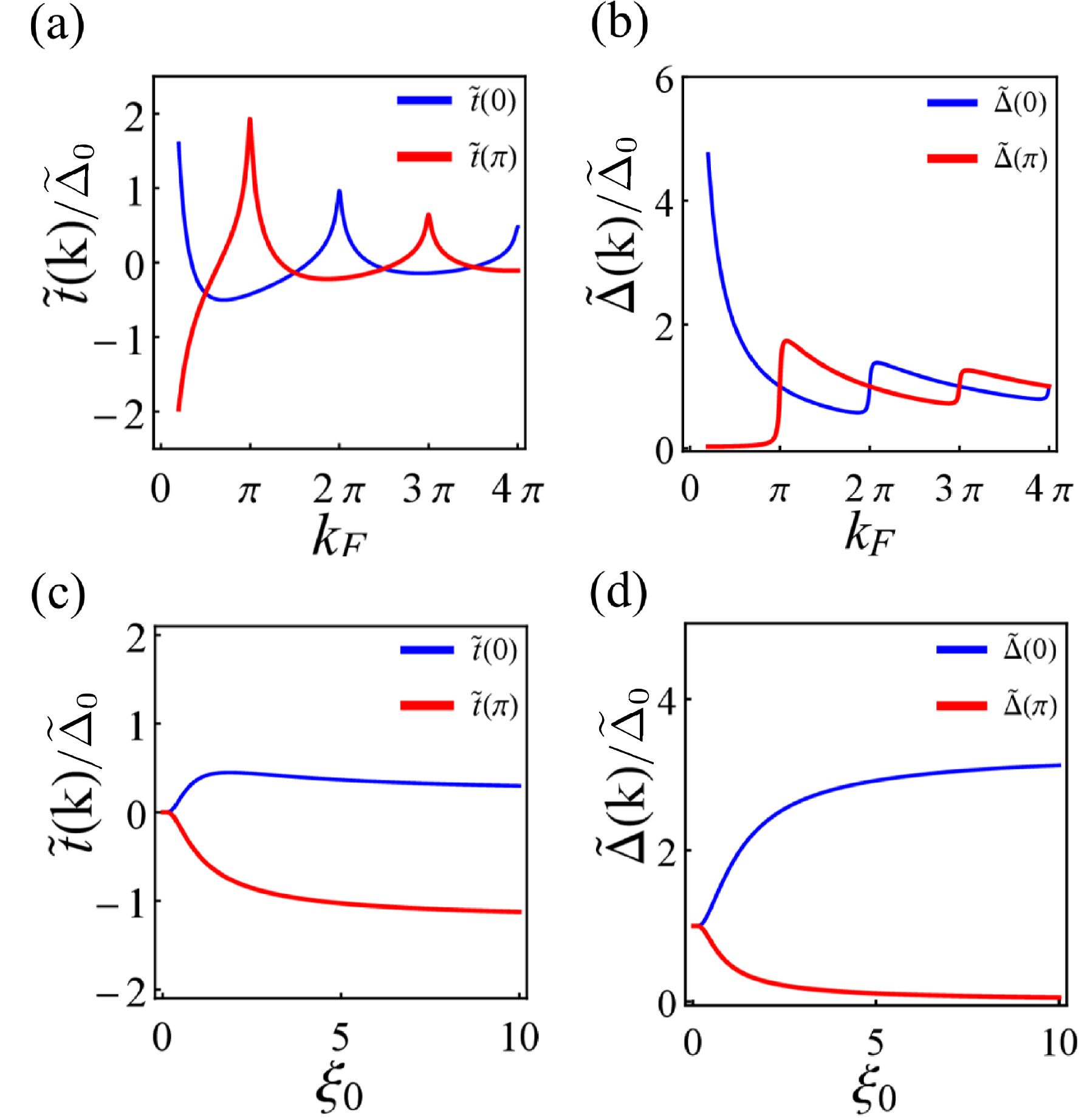}
	\caption{$\widetilde{t}(k)$ and $ \widetilde{\Delta}(k) $ as functions of $ k_F $ with $ \xi_{0}=10 $ [(a) and (b)] and of $ \xi_{0} $ with $k_F=0.3\pi$
		[(c) and (d)] for $ k=0$ (blue) and $k=\pi $ (red).}\label{deltt}
\end{figure}
For discussions on the topological phase diagram, we plot $ \widetilde{t}(k=0,\pi) $ and $ \widetilde{\Delta}(k=0,\pi) $ with $\xi_0=10$ as functions of $k_F$ in Figs.~\ref{deltt}(a) and (b), respectively.
All of them exhibit damped oscillation with increasing $k_{F}$. When $k_{F}<\pi$, $\widetilde{\Delta}(\pi)$ is much smaller than $\widetilde{\Delta}(0)$. In Figs.~\ref{deltt}(c) and (d), $ \widetilde{t}(k=0,\pi) $ and $ \widetilde{\Delta}(k=0,\pi) $ are plotted as functions of $\xi_0$, respectively, with $k_F =0.3\pi$. Each of them tends to be a constant value for $ \xi_{0} >10 $. Note that $ \widetilde{\Delta}(\pi) $ approaches zero for $ \xi_{0}$ large enough, indicating a quite small pairing potential at $ k=\pi $, in contrast to the case of the strong coupling limit.

\section{The Derivation of Topological Invariant}
In this section, we detail the derivation of the Berry phase formula, Eq.~(12), in the main text. The BdG Hamiltonian, Eq.~(10), of the nanowire can be block anti-diagonalized by a unitary transformation given by $U=e^{-\frac{\pi}{4}i\tau_{1}\otimes\sigma_{2}}$, namely, that
\begin{equation}
Uh(k)U^{\dagger}=\begin{pmatrix}
O&Q(k)\\Q^\dagger(k)&O
\end{pmatrix},
\end{equation}
where 
\begin{equation}
Q(k)=\left(\widetilde{\Delta}(k)+2i\lambda_{R}\sin k\right)-i\left(2t\cos k-\bar{\epsilon}+\widetilde{t}(k)\right)\sigma_{2}-h_Z\sigma_{3}.
\end{equation}
For the purpose of topological analysis, we consider the flattened Hamiltonian $\tilde{h}=\sum_\alpha|+,\alpha,k\rangle\langle+,\alpha,k|-\sum_\alpha|-,\alpha,k\rangle\langle-,\alpha,k|$, where $|\pm,\alpha\rangle$ are conduction and valence eigenstates of $h(k)$, respectively, with $\alpha=1,2$. Accordingly, the upper-right block of $Uh(k)U^{\dagger}$ is a unitary matrix $\hat{Q}(k)$ with
\begin{equation}
\hat{Q}^\dagger\hat{Q}=1_2. \label{unitary}
\end{equation}
Then, the valence eigenstates  can be given by
\begin{equation}
|-,\alpha,k\rangle=\frac{1}{\sqrt{2}}\begin{pmatrix}
-v_\alpha\\
\hat{Q}^\dagger(k)v_\alpha
\end{pmatrix},
\end{equation}
where $v_\alpha$ are an orthonormal basis of $\mathbb{C}^2$ independent of $k$. Hence, the Berry connection of valence bands is given by
\begin{equation}
\begin{split}
a(k)&=i\sum_{\alpha}\langle-,\alpha,k|\partial_k|-,\alpha,k\rangle\\
&=\frac{i}{2}\mathrm{tr}\hat{Q}\partial_k\hat{Q}^\dagger=-\frac{i}{2}\mathrm{tr}\hat{Q}^\dagger\partial_k\hat{Q}\\
&=-\frac{i}{2}\partial_k\mathrm{tr}\ln\hat{Q}=-\frac{i}{2}\partial_k\ln \mathrm{Det}\hat{Q}.
\end{split}
\end{equation}
In the second equality of the second line, we have used $\hat{Q}^\dagger\partial_k\hat{Q}=-(\partial_k\hat{Q}^\dagger)\hat{Q}$, which can be derived from Eq.~\eqref{unitary}. In the second equality of the third line, we recalled that $\ln \mathrm{Det}A=\mathrm{tr}\ln A$ for any matrix $A$. The determinant of $\hat{Q}$ is related to that of $Q$ by
\begin{equation}
\mathrm{Det}\hat{Q}(k)=\frac{1}{d(k)}\mathrm{Det}Q(k),
\end{equation}
where $d(k)$ is a real and definitely positive function of $k$. 
Thus, the Berry phase in unit of $\pi$ over the first Brillouin zone is
\begin{equation}
\label{winding}
W=\oint \frac{dk}{\pi}~a(k)=\oint \frac{dk}{2\pi i}~\partial _{k}\ln Z(k) \mod 2,
\end{equation}
where 
\begin{equation}
\label{Z}
Z(k)=(\widetilde{\Delta}(k)+2i\lambda_{R}\sin k)^2-h_{Z}^{2}+[2t\cos k-\bar{\epsilon}+\widetilde{t}(k)]^{2}. 
\end{equation}
Note that the Berry phase can be changed by an even integer under a gauge transformation, and $\partial_k \ln d(k)$ vanishes under the closed integration, since $d(k)$ is a real function.

Let $ Z(k)=\left|Z(k) \right|e^{i\theta(k)}  $ with $ \theta(k)=\arg Z(k) $. We can write Eq.~\eqref{winding} as
\begin{equation}
\label{wind}
\begin{split}
W=\oint\frac{dk}{2\pi i}~\left(\partial _{k}\ln \left| Z(k)\right| +i\partial_{k}\theta(k)\right)\mod 2.
\end{split}
\end{equation}
Because of particle-hole symmetry, we have the identity,
\begin{equation}
Z(k)=Z^*(-k),\label{zk}
\end{equation}
which can also be directly verified from Eq.~\eqref{Z}.
Since the nonzero real function $ \left| Z(k)\right| $ is  periodic over the first Brillouin zone, the first term in Eq.~\eqref{wind} vanishes for the closed path. For $\theta(k)$, equation~\eqref{zk} implies
\begin{equation}
\theta(k)=-\theta(-k)\ \mathrm{mod}\ 2\pi,
\end{equation}
and
\begin{equation}
\theta(0),~\theta(\pi)\in \{n\pi|~n\in\mathbb{Z}\}.
\end{equation}  
Thus,
\begin{equation}
\label{wind1}
\begin{split}
W&=\int_{0}^{\pi}\frac{dk}{\pi}\partial_{k}\theta(k)\ \mathrm{mod}\ 2\\
&=\frac{1}{\pi}\left[\theta(\pi)-\theta(0)\right]\ \mathrm{mod}\ 2.
\end{split}
\end{equation}
which is equivalent to the formula of Eq.~(12) in the main text.

\section{Energy Spectrum and Boundary States}

\begin{figure}
	\centering
	\includegraphics[scale=0.6]{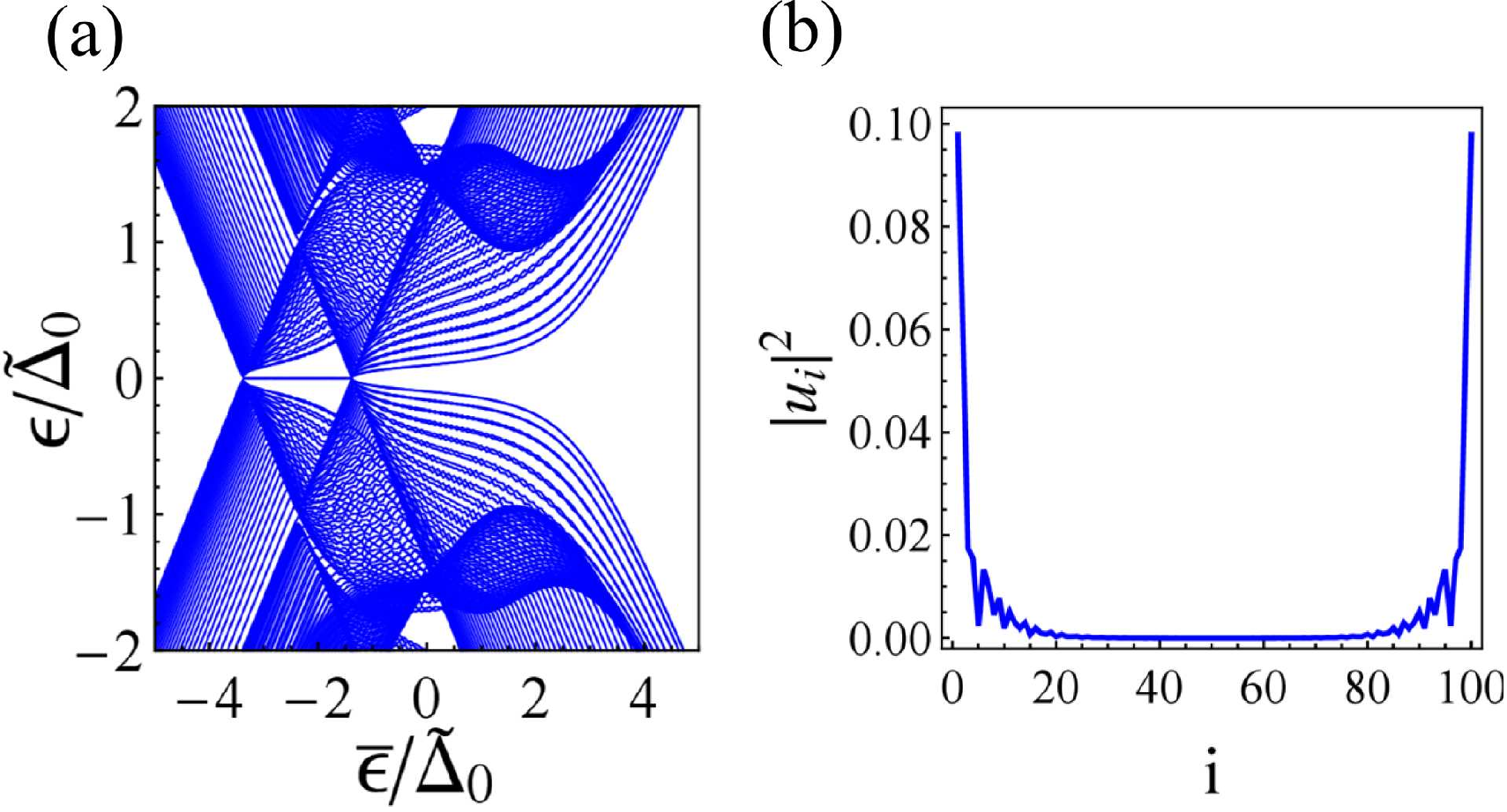}
	\caption{(a) Energy spectrum as a function of $\bar{\epsilon}$ under open boundary conditions with 100 lattice sites, $h_{Z}=1$, $\alpha_{R}=0.8$,  and the same $t,\widetilde{\Delta}_{0},\xi_{0},k_{F}$ as those in Fig.~\ref{diag}(c). (b) The spatial distribution of MFs for the $\bar{\epsilon}=2$ case in (a).}\label{fig3}
\end{figure}

To demonstrate the bulk-end correspondence, we numerically calculate the energy spectrum as a function of $\bar{\epsilon}$ under the Dirichlet boundary condition with 100 lattice sites, where we choose $h_{Z}=1$, $\lambda_{R}=0.8$, and the same values of $t,\widetilde{\Delta}_{0},\xi_{0},k_{F}$ as those in Fig.~1(c) of the main text.  By tuning $\bar{\epsilon}$, the appearance of Majorana zero modes is consistent with the phase diagram of Fig.~1(c) in the revised manuscript. These Majorana zero modes locate at the ends of the nanowire, which is clear from the probability density of their wave functions in real space, as plotted in Fig.~\ref{fig3}(b) with $\bar{\epsilon}=-2$.

\end{document}